\documentclass[aip,apl,graphicx,twocolumn,reprint]{revtex4}

\usepackage{graphicx} 				

\providecommand{\DIFdel}[1]{}

\begin{document}


\title{Submicrosecond-timescale readout of carbon nanotube mechanical motion} 



\author{H. B. Meerwaldt}
\author{S. R. Johnston}
\author{H. S. J. van der Zant}
\author{G. A. Steele}
\affiliation{Kavli Institute of Nanoscience, Delft University of Technology, Lorentzweg 1, 2628 CJ Delft, the Netherlands}
`


\begin{abstract}
We report fast readout of the motion of a carbon nanotube mechanical resonator. A close-proximity high electron mobility transistor amplifier is used to increase the bandwidth of the measurement of nanotube displacements from the kHz to the MHz regime. Using an electrical detection scheme with the nanotube acting as a mixer, we detect the amplitude of its mechanical motion at room temperature with an intermediate frequency of 6 MHz and a timeconstant of 780 ns, both up to five orders of magnitude faster than achieved before. The transient response of the mechanical motion indicates a ring-down time faster than our enhanced time resolution, placing an upper bound on the contribution of energy relaxation processes to the room temperature mechanical quality factor.

\end{abstract}


\maketitle 

Suspended carbon nanotubes (CNTs) are bottom-up high-aspect ratio mechanical resonators, which are excellent mass sensors \cite{Chaste2012}, couple strongly to single-electron tunneling \cite{Lassagne2009,Steele2009,Meerwaldt2012a}, and exhibit nonlinear damping \cite{Eichler2011a} and modal interaction \cite{Eichler2012,Castellanos-Gomez2012}.  A significant challenge for CNT nanoelectromechanical systems (NEMS) is the high electrical impedance of the CNT, which makes it difficult to read out its mechanical motion electrically at the resonance frequency of 100 MHz or higher. 

One approach to measure high-frequency mechanical resonances in such high-impedance devices involves decreasing the frequency at which the mechanical signal is detected by using the CNT itself as a mixer. The signal can be read out straightforwardly by down-mixing it to several kHz through two-source mixing techniques \cite{Knobel2003,Sazonova2004, Peng2006, Witkamp2006}, through frequency-modulation mixing \cite{Gouttenoire2010}, or by rectifying it to dc \cite{Huettel2009}. While such solutions solve part of the problem by shifting the signal down to low frequencies, a disadvantage is that the measurement bandwidth, i.e. the time resolution with which the mechanical amplitude can be followed, is still restricted by the high electrical impedance. Typically, the time resolution in such experiments is limited to $\sim$100 ms. 

A second approach involves connecting the high-impedance device to a low-impedance amplifier \cite{Xu2010, Abhilash2012}. Doing so, a high bandwidth is achieved, but the signal is decreased by the ratio of the impedances, typically a factor of 1000. Using this approach, long averaging times are required to detect the mechanical signal. In this Letter, we achieve a high readout bandwidth without sacrificing the amplitude of the signal by placing a high electron mobility transistor (HEMT) amplifier with a high input impedance \cite{Vink2007} only millimeters from the device. Doing so, we enhance the measurement bandwidth of the CNT mechanical mixer by five orders of magnitude, achieving single-shot detection of the CNT motion at submicrosecond timescales.

\begin{figure}[hb]
 \includegraphics{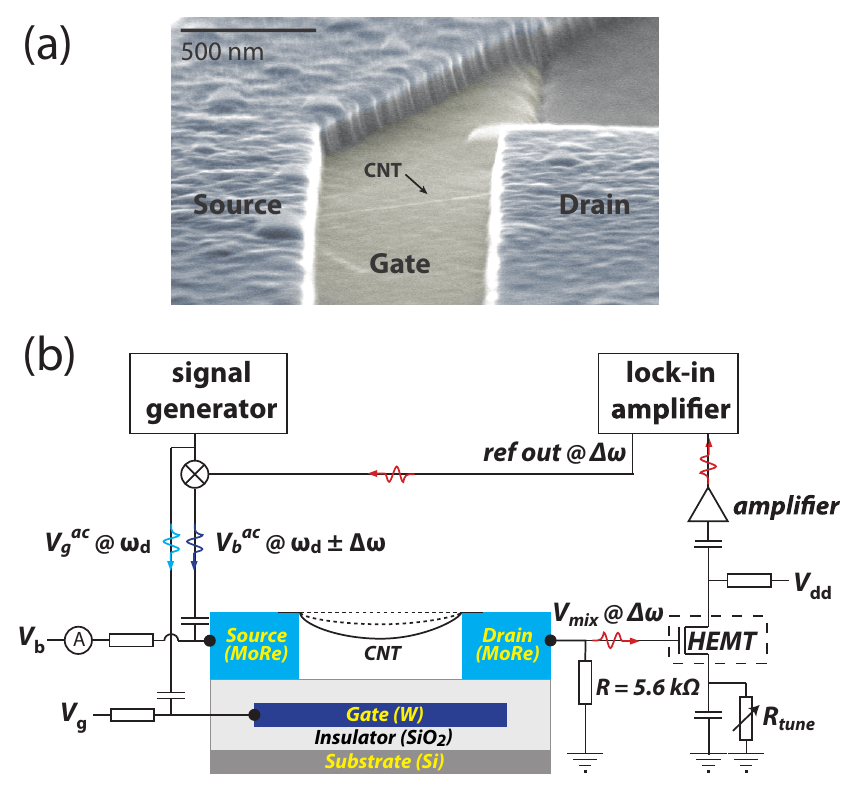}
 \caption{(a) A high-angle colorized SEM image of a typical device, showing a CNT suspended between the source and drain contacts. (b) Measurement setup. Excitation is performed using a signal generator, a lock-in amplifier, and a mixer, supplying signals at $\omega_d$ and $\omega_d \pm \Delta\omega$.  The device consists of a CNT suspended between two MoRe electrodes and a tungsten local gate buried in SiO$_2$. The CNT mixes the two signals mechanically and electrically down to frequency $\Delta\omega$.  The HEMT circuit measures the mixing current via the voltage it generates over the 5.6 k$\Omega$ resistor. The voltage on the gate of the HEMT amplifier modulates the current supplied by $V_{dd}$, which is subsequently amplified and measured by the lock-in amplifier. }
\label{fig:setupplusHEMT}
 \end{figure}

A scanning electron microscope (SEM) image of a typical device is shown in figure \ref{fig:setupplusHEMT}(a). Devices are made on an oxidized high resistivity silicon wafer with a local gate to excite the mechanical motion and minimize capacitive crosstalk to the source and drain electron. Fabrication of the CNT resonators takes place as follows. First, 50 nm of tungsten is deposited onto the substrate by sputtering. Local gates are defined in the tungsten with SF$_6$/He dry etching, after which they are covered with 200 nm of silicon oxide using plasma-enhanced chemical vapor deposition. Next, a 35 nm layer of molybdenum and a 35 nm layer of rhenium are deposited using sputtering. Using a four-layer etch mask\cite{Meerwaldt2012a}, the electrodes are defined by dry etching, while at the same time forming a self-aligned trench in the silicon oxide. The local gate bond pads are uncovered with buffered hydrofluoric wet etching and a layers of chromium, platinum, and silicon deposited as metallization. In the final step, catalyst sites are defined and CNTs are grown across the structures using chemical vapor deposition. For CNT growth, an alumina-supported Fe/Mo catalyst is used that produces a high-yield of single-wall CNTs with diameters in the range of 1-3 nm.

\begin{figure}[b]
 \includegraphics{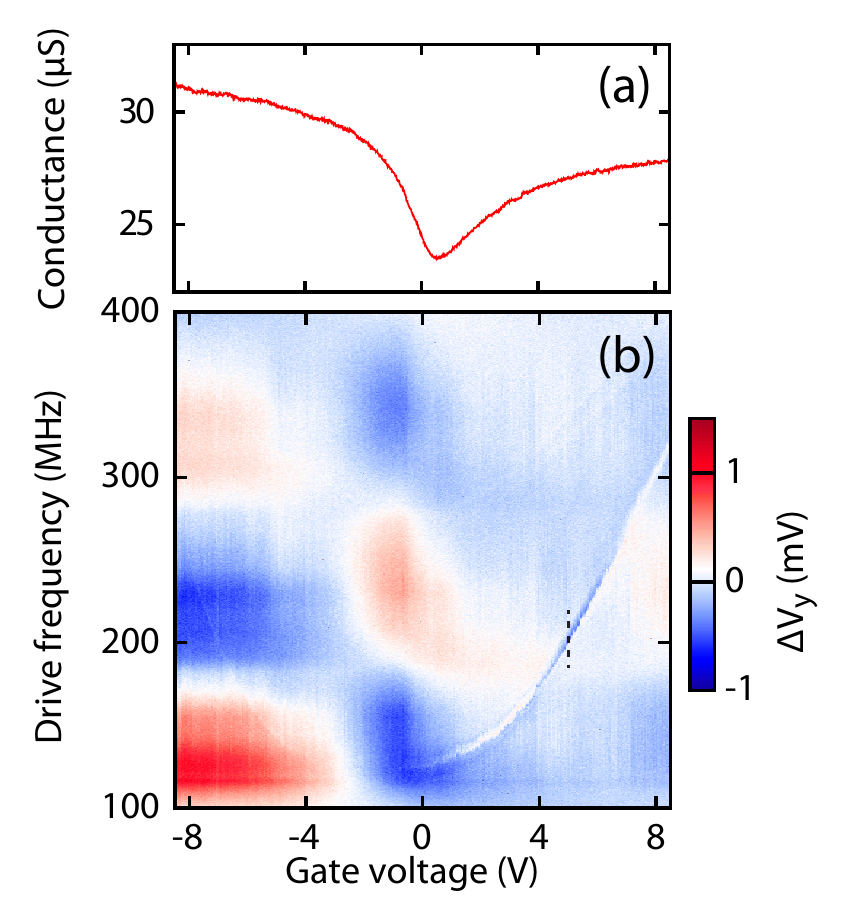}%
 \caption{A carbon nanotube electromechanical resonator: (a) Conductance of the CNT in series with two 5.6 k$\Omega$ resistors as a function of gate voltage, showing a small-bandgap CNT. (b) Y-component of the mixing voltage at $\Delta\omega/2\pi=6$ MHz and $\tau=$ 1 $\mu$s as a function of the gate voltage and drive frequency, $V_G^{ac} = -13.6$ dBm (66 mV), $V_{sd}^{ac} =$ -19 dBm (33 mV). Off-resonant current has been subtracted for each frequency. Resonances of three mechanical modes, one clear fundamental mode and two fainter higher order modes, are visible as lines shifting upwards with positive or negative gate voltage. The dashed line indicates the gate voltage and frequencies corresponding to the linecut displayed in Figure \ref{fig:howfast}(b).}
\label{fig:mechres}
 \end{figure}

To perform the high-bandwidth readout of the CNT mechanical resonator, a two-source mixing measurement scheme is used in room-temperature vacuum, as shown in Figure \ref{fig:setupplusHEMT}(b). The excitation is provided by a Rohde \& Schwartz SMB100A signal generator. Measurements are performed using a Zurich Instruments HF2LI high-frequency lock-in amplifier which is able to operate with a time constant down to 780 ns. The ac gate voltage at $\omega_d$ is supplied by the signal generator. The ac  voltage applied to the source contact at $\omega_d \pm \Delta \omega$ is generated by mixing the signal generator's output with the reference signal of the lock-in amplifier at $\Delta \omega$. The ac voltages at the source and gate are mixed by the CNT, both mechanically and electrically, producing a mixing current at frequency $\Delta \omega$. The gate of the HEMT amplifier is connected by a short PCB trace to the drain of the CNT. The mixing current from CNT is converted to a voltage at the gate of the HEMT amplifier using a 5.6 k$\Omega$ readout resistor. This voltage modulates the current supplied by $V_{dd}$, which is subsequently amplified and measured by the lock-in amplifier.  In the experiments here, the HEMT is used predominantly as an impedance matching amplifier with relatively low gain, and can therefore be operated at a cryogenically compatible power consumption of 120 $\mu$W.

In Figure \ref{fig:mechres}, we characterize the electrical and mechanical properties of the CNT device. Figure \ref{fig:mechres}(a) shows the conductance through the CNT resonator at room temperature as a function of gate voltage. The conductance is measured by applying a dc bias voltage of 10 mV and measuring the current. The v-shaped conductance indicates an small-bandgap CNT. Figure \ref{fig:mechres}(b) shows the Y-component of the mixing voltage as a function of gate voltage and drive frequency. The Y-quadrature of the lockin signal is plotted as it shows the highest contrast. For clarity, the background current for each drive frequency is subtracted. In figure \ref{fig:mechres}(b), a mechanical resonance can be seen as a sharp line in the plot which shows the characteristic gate dependence of the fundamental mode \cite{Sazonova2004,Witkamp2006}. For small deviations from the minimum, the mechanical frequency is relatively flat, with a gate voltage dependence that appears flatter than quadratic, indicating that our CNT does not have any slack \cite{Sazonova2004,Witkamp2006}, consistent with the SEM image of a typical device (Fig.\ \ref{fig:setupplusHEMT}(a)). Two faint lines at higher frequency correspond to higher modes. Although the mechanical response, shown in Figure 3(b), is non-linear at the drive powers used, we can estimate a lower bound on the quality factor $Q > 20$ from the full-width at half-maximum of the resonance.

\begin{figure}[hb]
\includegraphics[scale=0.99]{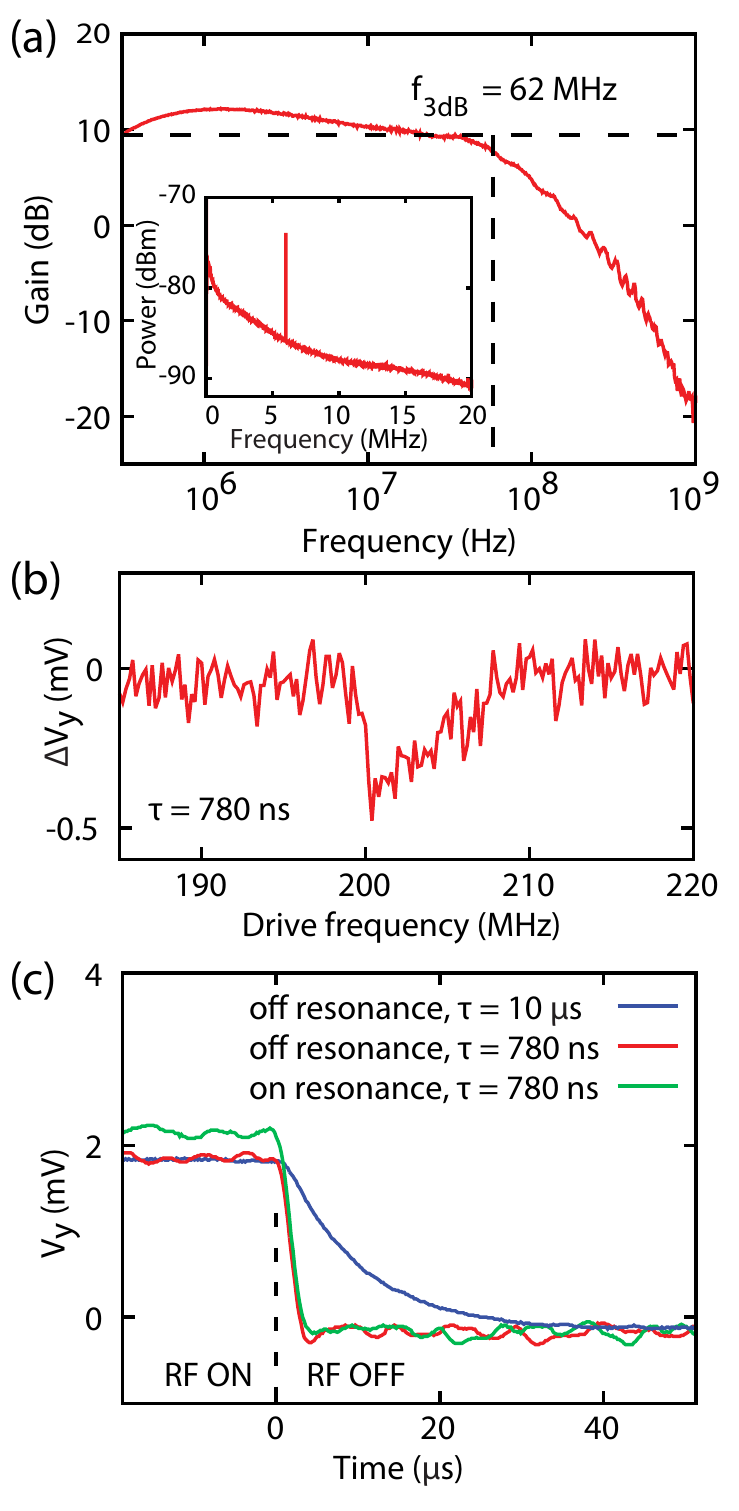}%
\caption{Fast detection of nanotube mixing signals: (a) Gain of the HEMT amplifier, showing a bandwidth of $f_{3dB}=62$ MHz. Inset: Output spectrum of the HEMT amplifier with the carbon nanotube acting as a high-IF bandwidth mixer, showing a peak at $\Delta\omega/2\pi=6$ MHz. (b) Mechanical resonance at $V_g = 4.95$ V, measured with $\Delta\omega/2\pi=6$ MHz and $\tau=$ 780 ns. The mechanical signals is clearly distinguishable from the noise with averaging times on the sub-microsecond timescale. (c) Single-shot measurement of the decay of the mixing signal at $\Delta\omega/2\pi = 6$ MHz, showing the decay from an off-resonance drive voltage with a timeconstant of $\tau=10$ $\mu$s (blue) and $\tau = 780$ ns (red), and with a drive voltage tuned to the mechanical resonance with $\tau$ = 780 ns (green). }
\label{fig:howfast}
\end{figure}

To determine the bandwidth of the HEMT amplifier, a transmission measurement is performed at zero gate voltage. This is done by applying a RF signal to the source of the CNT and measuring the signal out of the HEMT amplifier. Figure \ref{fig:howfast}(a) shows the gain of the circuit. The 24 dB amplification of the second amplifier, and the attenuation due to the voltage divider of the CNT's resistance with the readout resistance are accounted for. Qualitatively, the behavior of the gain can be understood using a first-order RC model of the CNT resistor and the drain's parasitic capacitance, which causes the roll-off above $f_{3dB}$. Above 24 MHz, a gain of 9.3 dB is visible and a bandwidth of $f_{3dB}=62$ MHz is determined. From this, the parasitic capacitance is calculated to be 0.5 pF, slightly higher than the input capacitance of the Avago ATF-35143 HEMT amplifier \cite{Vink2008} of 0.4 pF. To demonstrate a significant increase in readout bandwidth, the inset of Figure \ref{fig:howfast}(a) shows the output spectrum of the HEMT, while applying ac voltages at 200 MHz on the gate and at 194 MHz and 206 MHz on the source of the CNT. The down-mixed signal appears as a peak in the output spectrum at 6 MHz, demonstrating an increase in the intermediate-frequency (IF) bandwidth of a CNT mixer by nearly three orders of magnitude compared to previous CNT mechanical mixing experiments.

In Figure \ref{fig:howfast}(b), we demonstrate readout of the CNT's mechanical motion on a submicrosecond timescale. The frequency trace ($V_g^{dc} = 4.95$ V, indicated in Fig. \ref{fig:mechres}(b), $\Delta \omega/2\pi=$ 6 MHz, $\tau= 780$ ns) shows the mechanical resonance clearly as a dip in the lockin Y-quadrature signal of $V_Y^{mech}=330$ $\mu$V. The amplitude of the mechanical oscillation is estimated by comparing \cite{Witkamp2006} the mechanical contribution of the mixing signal, $V_Y^{mech}$, to the electrical contribution, $V_Y^{elec}=5$ mV: 
\begin{equation}
u = \frac{V_Y^{mech}}{V_Y^{elec}}\frac{V_g^{ac}}{V_g^{dc}}h_0 \ln(2h_0/r).
\label{eq:}
\end{equation}
Using the distance between CNT and gate $h_0 = 200$ nm, a typical single-walled CNT's radius $r = 1-3$ nm, and the excitation voltage on the gate $V_g^{ac} = 66$ mV, the amplitude of motion is estimated to be 1.05-0.86 nm. The lineshape of the mechanical resonance depends on the phase difference between the mechanical and the electrical mixing current. The sharp edge in Fig. \ref{fig:howfast}(b) suggests nonlinear oscillation, which is expected at an amplitude comparable to the CNT's radius. The total active measurement time for Figure \ref{fig:howfast}(b) is 140 $\mu$s, allowing determination of the resonance lineshape by a full frequency sweep with submillisecond time resolution. 

In Figure \ref{fig:howfast}(c), we use the high readout bandwidth of our technique to explore the transient response of the mechanical motion of carbon nanotube resonators at room temperature.  While carbon nanotubes exhibit large quality factors at low temperatures, $Q\sim10^5$ \cite{Huettel2009}, the quality factor observed at room temperature is very low \cite{Sazonova2004}, $Q\sim100$. A possible origin of such observed quality factors is spectral broadening from thermal motion of other mechanical modes \cite{Barnard2012}, an effect that is strong in carbon nanotubes due to their strong mode coupling\cite{Eichler2012,Castellanos-Gomez2012}. A signature of such spectral broadening would be a difference between the mechanical quality factor $Q_{s}$ measured from the spectral linewidth of the resonance and the quality factor $Q_{r}$ measured from a ring-down experiment.

To study the time-domain response of the nanotube with our high frequency readout, we apply a pulsed a.c. drive voltage to the gate and record the transient response of the electrical and mechanical mixing currents.  The drive signal $V_g$ is pulse modulated with a radio-frequency switch, and we record the Y-component of the mixing signal from the analog output of the lockin amplifier with an oscilloscope. Figure \ref{fig:howfast}c shows a single-shot measurement of the decay of the off-resonance mixing signal at a timeconstant of 10 $\mu$s (blue) and 780 ns (red). As expected from the large bandwidth of the CNT-HEMT amplifier circuit, the decay of the electrical mixing current is dominated by the lockin amplifier's timeconstant. The green line shows the recorded transient for a drive signal tuned to a mechanical resonance with $\tau$ = 780 ns. The mechanical component of the mixing signal can be seen by the difference in height between the green and the red traces while the RF is turned on. As can be seen in the figure, the mixing singal on-resonance decays just as fast as the the electrical signal off resonance, indicating that the mechanical ring-down time of the CNT at room temperatures is faster than 780 ns. This is also found at other gate voltages (not shown). These observations give an upper bound on the energy relaxation contribution to the room-temperature quality factor of $Q_r < \omega_0\tau/2 = 490$.

In addition to such transient experiments to explore mechanical ring-down, there are also several other applications where the high measurement bandwidth of our technique could be useful. When the mechanical resonator is driven to large amplitudes, nonlinear dynamics can be observed while the resonator decays into one of the two bistable states \cite{Unterreithmeier2010}, potentially in the quantum regime \cite{Dykman1988, Katz2007}. The high-bandwidth readout allows measuring with a large $\Delta \omega$, outside of the mechanical bandwidth. Undriven motion such as thermal motion \cite{Stapfner2012}, zero-point motion, and self-sustained oscillation through single-electron tunneling \cite{Steele2009,Usmani2007} can now be investigated because the mechanical motion can be probed without driving it. Finally, mass sensing experiments \cite{Chaste2012} can be performed with significantly faster time resolution, opening up the possibility of probing phenomena such as the diffusion of adsorbates on the mechanical resonator.

In summary, we have performed high-frequency mixing of a CNT mechanical resonator using a close-proximity high-impedance HEMT amplifier. The CNT-HEMT system has a bandwidth of 62 MHz. We show a mechanical resonance at a $\Delta \omega/2\pi$ of 6 MHz and a timeconstant of 780 ns, up to five orders of magnitudes faster than reported previously. The transient response of the mechanical signal gives an upper bound to the room-temperature quality factor due to energy relaxation of $Q_r < 490$. Future work will focus on mechanical transients at cryogenic temperatures, which are expected to have longer decay times due to the higher quality factor of the CNT resonator, and exploring the potential of this high-bandwidth technique in other carbon NEMS applications.

We thank Raymond Schouten and Vibhor Singh for helpful discussions and advice, and Ben Schneider for help with fabrication of the devices. This work is supported by the Dutch Foundation for Fundamental Research on Matter (FOM), the Netherlands Organisation for Scientific Research (NWO), and the Europeon Union (FP7) through project RODIN.


%
%

%




\end{document}